\documentclass[multphys,vecphys]{svmult}

\usepackage{makeidx}         % allows index generation
\usepackage{graphicx}        % standard LaTeX graphics tool
\usepackage{multicol}        % used for the two-column index
\usepackage[bottom]{footmisc}% places footnotes at page bottom

\makeindex             % used for the subject index
\begin{document}

\title*{Revisiting VLT/UVES constraints on a varying fine-structure constant}
\titlerunning{Revisiting VLT/UVES constraints on varying $\alpha$}
\author{M.~T.~Murphy\inst{1} \and
J.~K.~Webb\inst{2} \and V.~V.~Flambaum\inst{2}}
\authorrunning{Murphy et al.}
\institute{Institute of Astronomy, University of Cambridge, Madingley Road, Cambridge CB3\ 0HA, UK;
\texttt{mim@ast.cam.ac.uk}
\and School of Physics, University of New South Wales, Sydney, NSW 2052, Australia; \texttt{jkw@phys.unsw.edu.au, flambaum@phys.unsw.edu.au}}
\maketitle

\begin{abstract}
  Current analyses of VLT/UVES quasar spectra disagree with the
  Keck/HIRES evidence for a varying fine-structure constant, $\alpha$.
  To investigate this we introduce a simple method for calculating the
  minimum possible uncertainty on $\Delta\alpha/\alpha$ for a given
  quasar absorber. For many absorbers in Chand et al.~(2004) and for the
  single-absorber constraint of Levshakov et al.~(2006) the quoted
  uncertainties are smaller than the minimum allowed by the UVES data.
  Failure of this basic consistency test prevents reliable comparison
  of the UVES and HIRES results.
\end{abstract}

\vspace{-10mm}\section{Introduction}\vspace{-3mm}

The `many-multiplet' (MM) method is the most precise technique for
constraining cosmological changes in the fine-structure constant,
$\alpha$, from QSO absorption spectra \cite{WebbJ_99a}. We have
previously described self-consistent MM evidence from 143 Keck/HIRES
absorbers for a smaller $\alpha$ over the redshift range $0.2\le
z_{\rm abs}\le4.2$ at the fractional level
$\Delta\alpha/\alpha=(-0.57\pm0.11)\times10^{-5}$ \cite{MurphyM_04a}.
Clearly, independent analyses of spectra from different spectrographs
are desirable to refute/confirm this. First attempts with VLT/UVES
spectra, e.g.~\cite{ChandH_04a,LevshakovS_06b}, generally found null
results with quoted uncertainties $<0.1\times10^{-5}$. To investigate
these claims, we introduce a simple method for calculating the minimum
possible uncertainty on $\Delta\alpha/\alpha$ from a given absorption
system.

\vspace{-3mm}\section{A simple measure of the limiting precision on
  $\Delta\alpha/\alpha$}\vspace{-3mm}

The velocity shift, $v_j$, of transition $j$ due to a small relative
variation in $\alpha$, $\Delta\alpha/\alpha\ll 1$, is determined by
the $q$-coefficient for that transition,
\begin{equation}\label{MurEq:da}
\omega_{z,j} \equiv \omega_{0,j} + q_j\left[\left(\alpha_z/\alpha_0\right)^2-1\right]\,\hspace{1em}\Rightarrow\hspace{1em}\frac{v_j}{c} \approx -2\frac{\Delta\alpha}{\alpha}\frac{q_j}{\omega_{0,j}}\,,
\end{equation}
where $\omega_{0,j}$ \& $\omega_{z,j}$ are the rest-frequencies in the
lab and at redshift $z$, $\alpha_0$ is the lab value of $\alpha$ and
$\alpha_z$ is the shifted value measured from an absorber at $z$. The
MM method is the comparison of measured velocity shifts from several
transitions (with different $q$-coefficients) to compute the best-fit
$\Delta\alpha/\alpha$. The linear equation (\ref{MurEq:da}) implies that
the error in $\Delta\alpha/\alpha$ is determined only by the
$q$-coefficients (assumed to have negligible errors) and the total
velocity uncertainty, integrated over the absorption profile, from
each transition, $\sigma_{{\rm v},j}$:
\begin{equation}\label{MurEq:lim}
\delta(\Delta\alpha/\alpha)_{\rm lim} = \sqrt{S/D}\,,
\end{equation}
for $S\equiv\sum_j\left(\frac{\sigma_{{\rm v},j}}{c}\right)^{-2}$ and
$D\equiv
S\sum_j\left(\frac{2q_j}{\omega_{0,j}}\right)^2\left(\frac{\sigma_{{\rm
        v},j}}{c}\right)^{-2} -
\left[\sum_j\frac{2q_j}{\omega_{0,j}}\left(\frac{\sigma_{{\rm
          v},j}}{c}\right)^{-2}\right]^2$.  This expression is just
the solution to a least-squares fit of $y=a+bx$ to data $(x_i,y_i)$,
with errors only on the $y_i$, where the intercept $a$ is also allowed
to vary; this mimics the real situation in fitting absorption lines
where the absorption redshift and $\Delta\alpha/\alpha$ must be
determined simultaneously.

The quantity $\sigma_{{\rm v},j}$ is commonly used in radial-velocity
searches for extra-solar planets, e.g.~\cite{BouchyF_01a}, but is not
normally useful in QSO absorption-line studies. Most metal-line QSO
absorption profiles display a complicated velocity structure and one
usually focuses on the properties of individual velocity components,
each of which is typically modelled by a Voigt profile. However, it is
important to realize that $\Delta\alpha/\alpha$ and its uncertainty
are integrated quantities determined by the entire absorption profile.
Thus, $\sigma_{{\rm v},j}$ should incorporate all the
velocity-centroiding information available from a given profile shape.
From a spectrum $F(i)$ with 1-$\sigma$ error array $\sigma_F(i)$, the
minimum possible velocity uncertainty contributed by pixel $i$ is
given by \cite{BouchyF_01a}
\begin{equation}\label{MurEq:sv_i}
\frac{\sigma_{\rm v}(i)}{c} = \frac{\sigma_F(i)}{\lambda(i)\,\left[\partial F(i)/\partial\lambda(i)\right]}\,.
\end{equation}
That is, a more precise velocity measurement is available from those
pixels where the flux has a large gradient and/or a small uncertainty.
This quantity can be used as an optimal weight, $W(i) \equiv
\left[\sigma_{\rm v}(i)/c\right]^{-2}$, to derive the total velocity
precision available from all pixels in a (portion of) spectrum,
\begin{equation}\label{MurEq:sv}
\textstyle\sigma_{\rm v} = c\left[\sum_i W(i)\right]^{-1/2}\,.
\end{equation}
For each transition in an absorber, $\sigma_{{\rm v},j}$ is calculated
from (\ref{MurEq:sv_i}) \& (\ref{MurEq:sv}); the only requirements are the
1-$\sigma$ error spectrum and the multi-component Voigt profile fit to
the transition's absorption profile. The latter allows the derivative
in (\ref{MurEq:sv_i}) to be calculated without the influence of noise.
Once $\sigma_{{\rm v},j}$ has been calculated for all transitions, the
uncertainty in $\Delta\alpha/\alpha$ follows from (\ref{MurEq:lim}).

\begin{figure}
\centering
\centerline{\hbox{
 \includegraphics[width=0.45\textwidth]{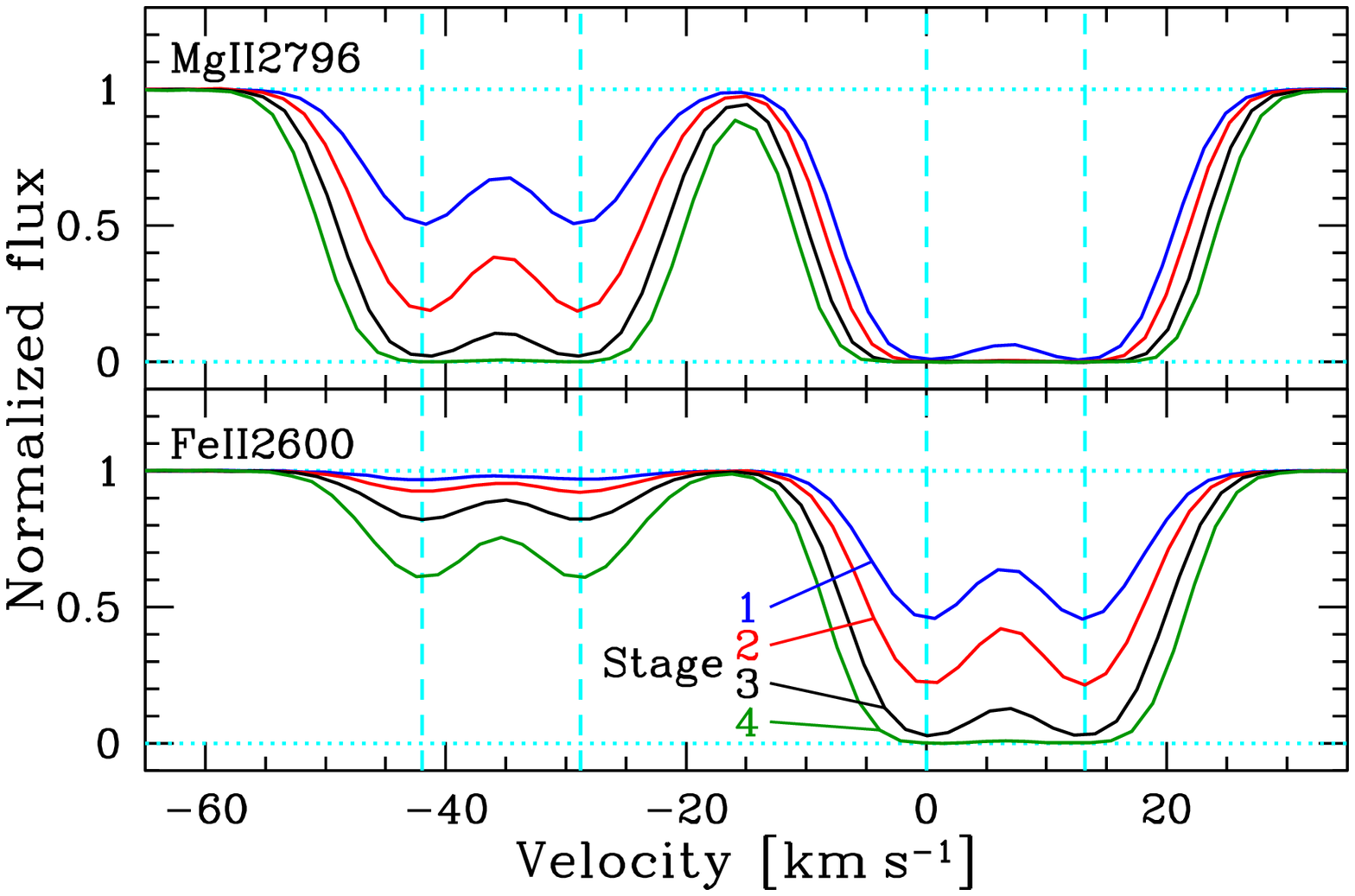}
 \hspace{0.01\textwidth}
 \includegraphics[width=0.47\textwidth]{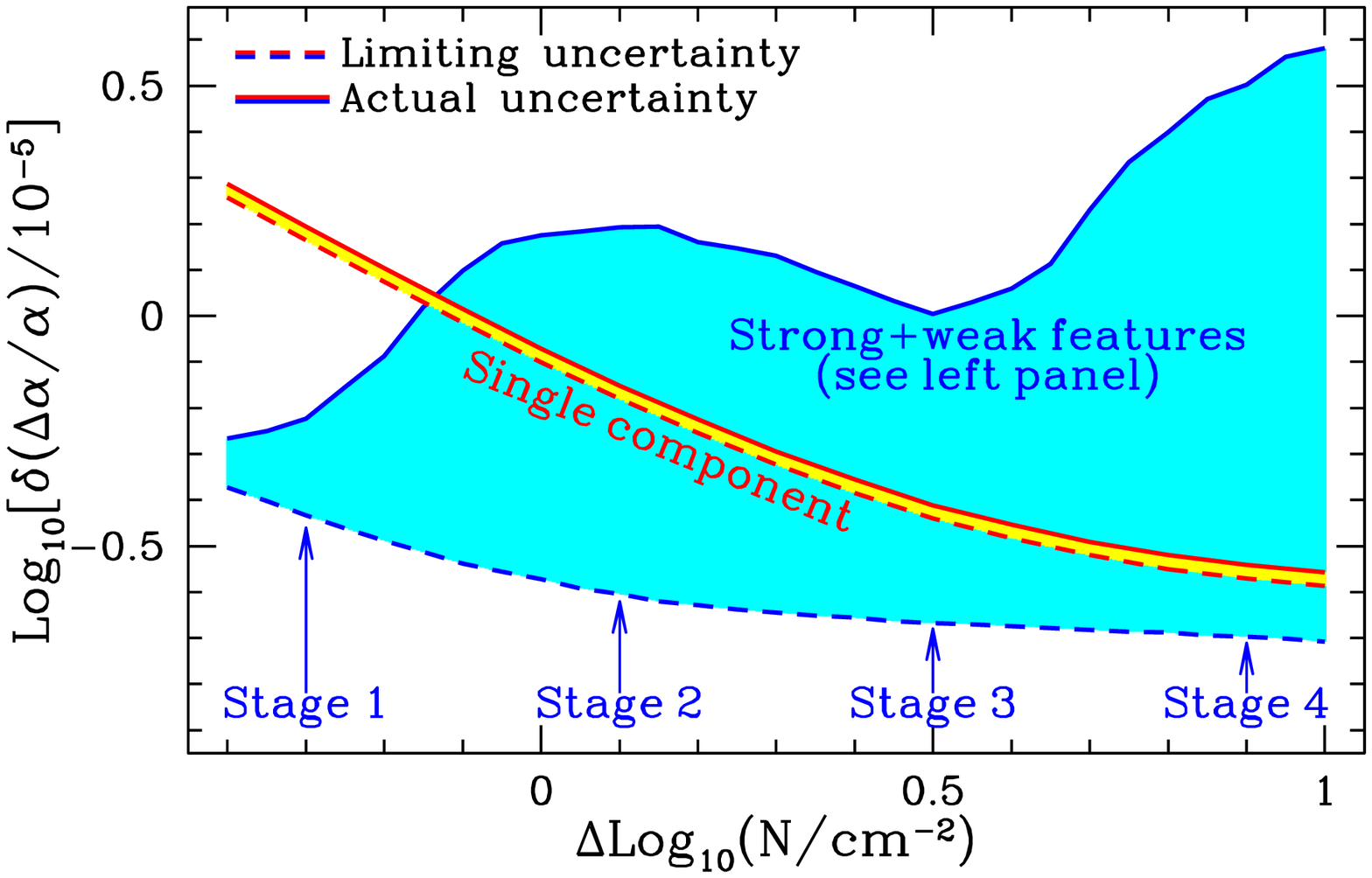}
}}
\vspace{-2mm}
\caption{\emph{Left}: Simulation of two transitions in a
  multi-component absorber. Labelled are distinct stages of
  differential saturation in the two main spectral features.
  \emph{Right}: The difference between the limiting precision,
  $\delta(\Delta\alpha/\alpha)_{\rm lim}$, and the actual precision
  (as derived by $\chi^2$-minimization analysis) varies strongly from
  stage to stage. In a single-component absorber the actual
  uncertainty tracks $\delta(\Delta\alpha/\alpha)_{\rm lim}$.}
\vspace{-3mm}
\label{MurFig:sim}
\end{figure}

It is important to realize that the uncertainty calculated with the
above method represents the absolute minimum possible 1-$\sigma$ error
on $\Delta\alpha/\alpha$; the real error -- as derived from a
simultaneous $\chi^2$-minimization of all parameters comprising the
Voigt profile fits to all transitions -- will always be larger than
$\delta(\Delta\alpha/\alpha)_{\rm lim}$ from (\ref{MurEq:lim}). The
main reason for this is that absorption systems usually have several
velocity components which have different optical depths in different
transitions. Equation (\ref{MurEq:lim}) assumes that the velocity
information integrated over all components in one transition can be
combined with the same integrated quantity from another transition to
yield an uncertainty on $\Delta\alpha/\alpha$. However, in a real
determination of $\Delta\alpha/\alpha$, each velocity component (or
group of components which define a sharp spectral feature) in one
transition is, effectively, compared with only the same component (or
group) in another transition. Fig.~\ref{MurFig:sim} illustrates this
point.

\vspace{-3mm}\section{Application to existing constraints on $\Delta\alpha/\alpha$}\vspace{-3mm}

We have calculated $\delta(\Delta\alpha/\alpha)_{\rm lim}$ for the
absorbers from the three independent data-sets which constitute the
strongest current constraints on $\Delta\alpha/\alpha$: (i) The 143
absorbers in our Keck/HIRES sample \cite{MurphyM_04a}; (ii) The 23
Mg/Fe{\sc \,ii} systems from UVES studied by \cite{ChandH_04a}; (iii)
The UVES exposures of the $z_{\rm abs}=1.1508$ absorber towards
HE\,0515$-$4414 studied by \cite{LevshakovS_06b}. For sample (ii), the
spectra were kindly provided by B.~Aracil who confirmed that they have
the same wavelength and flux arrays as those used in
\cite{ChandH_04a}.  The main difference is the error arrays: our error
array is generally a factor $\approx 1.4$ smaller than that used by
\cite{ChandH_04a} (H.~Chand, B.~Aracil, 2006, priv.~comm.). We have
confirmed this by digitizing the absorption profiles plotted in
\cite{ChandH_04a}.  Thus, $\delta(\Delta\alpha/\alpha)_{\rm lim}$
calculated using our spectra will be \emph{smaller} than the value
\cite{ChandH_04a} would derive. The continuum normalization will also
be slightly different, but this has negligible effects on the analysis
here. For sample (iii), the exposures were reduced using a modified
version of the UVES pipeline and their $S/N$ matches very well those
quoted by \cite{LevshakovS_06b}. For samples (ii) \& (iii) we use the
Voigt profile models published in \cite{ChandH_04a} \&
\cite{LevshakovS_06b} to calculate $\delta(\Delta\alpha/\alpha)_{\rm
  lim}$. For all samples, the atomic data for the transitions
(including $q$-coefficients) were the same as used by the original
authors. In practice, when applying (\ref{MurEq:sv_i}) and
(\ref{MurEq:sv}) we sub-divide the absorption profile of each
transition into $15{\rm \,km\,s}^{-1}$ chunks to mitigate the effects
illustrated in Fig.~\ref{MurFig:sim}. This provides a value of
$\delta(\Delta\alpha/\alpha)_{{\rm lim},k}$ for each chunk $k$. The
final value of $\delta(\Delta\alpha/\alpha)_{\rm lim}$ is simply
$\{\sum_k 1/[\delta(\Delta\alpha/\alpha)_{{\rm lim},k}]^2\}^{1/2}$; in
all cases this is $<1.4$ times the value obtained without
sub-divisions.

\begin{figure}
\centering
\includegraphics[width=0.60\textwidth]{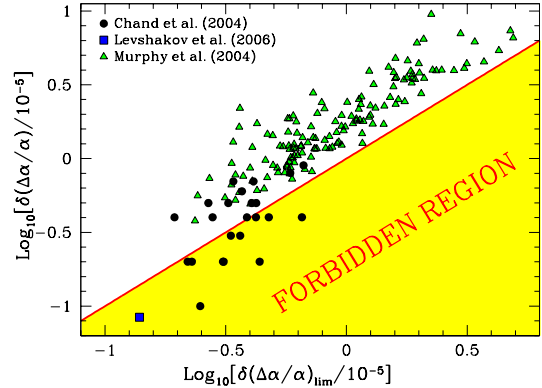}
\vspace{-2mm}
\caption{Quoted errors vs.~limiting precision for current samples.
  The Chand et al.~and Levshakov et al.~samples fail the basic
  requirement $\delta(\Delta\alpha/\alpha) >
  \delta(\Delta\alpha/\alpha)_{\rm lim}$.}
\vspace{-3mm}
\label{MurFig:lim}
\end{figure}

Figure \ref{MurFig:lim} shows the 1$\sigma$ error on
$\Delta\alpha/\alpha$ quoted by the original authors versus the
limiting precision, $\delta(\Delta\alpha/\alpha)_{\rm lim}$. The main
results are clear. Firstly, the 1$\sigma$ errors quoted for the HIRES
sample in \cite{MurphyM_04a} always exceed
$\delta(\Delta\alpha/\alpha)_{\rm lim}$, as expected if the former are
robustly estimated. Secondly, for at least 11 of their 23 absorbers,
Chand et al.~\cite{ChandH_04a} quote errors which are \emph{smaller}
than $\delta(\Delta\alpha/\alpha)_{\rm lim}$. Recall that since their
error arrays are larger than ours, 11 out of 23 is a conservative
estimate. Finally, the very small error quoted by
\cite{LevshakovS_06b} for HE\,0515$-$4414, $0.084\times10^{-5}$,
disagrees significantly with the limiting precision of
$0.14\times10^{-5}$. Thus, the (supposedly) strong current UVES
constraints on $\Delta\alpha/\alpha$ fail a basic consistency test
which not only challenges the precision claimed by \cite{ChandH_04a}
and \cite{LevshakovS_06b} but which must bring into question the
robustness and validity of their analysis and final
$\Delta\alpha/\alpha$ values.

\vspace{-3mm}\section{Conclusion}\vspace{-3mm}

We have introduced a very simple method of determining the limiting
precision on $\Delta\alpha/\alpha$ obtainable from a set of
transitions in a QSO absorption system,
$\delta(\Delta\alpha/\alpha)_{\rm lim}$. Only the 1-$\sigma$ error
spectrum and the model absorption profile (generally constructed from
Voigt profiles) are required to calculate
$\delta(\Delta\alpha/\alpha)_{\rm lim}$. The method simply equates the
total velocity information contained in the absorption profile of a
transition with the expected error on $\Delta\alpha/\alpha$ via that
transition's sensitivity to $\alpha$-variation. All absorption systems
in our HIRES data-set have quoted uncertainties in
$\Delta\alpha/\alpha$ which exceed $\delta(\Delta\alpha/\alpha)_{\rm
  lim}$, as expected.  However, the uncertainties on
$\Delta\alpha/\alpha$ currently quoted by \cite{ChandH_04a} and
\cite{LevshakovS_06b} from UVES spectra are, in many cases,
\emph{smaller} than $\delta(\Delta\alpha/\alpha)_{\rm lim}$. Clearly,
the UVES and HIRES results cannot be reliably compared until more
rigorous analyses of the UVES spectra are completed.

% BibTeX users please use
%\bibliographystyle{aveiro06}
%\bibliography{references}

%
% Non-BibTeX users please follow the syntax
% the syntax of "referenc.tex" for your own citations
%\input{referenc}
%%%%%%%%%%%%%%%%%%%%%%%%%%%%%%%%%%%%%%%%%%%%%%%%%%%%%%%%%%%%%%%%%%%%%% }

%%%%%%%%%%%%%%%%%%%%%%%%%%%%%%%%%%%%%%%%%%%%%%%%%%%%%%%%%%%%%%%%%%%%%%

\printindex
\end{document}